\documentclass[a4paper]{jpconf}
\usepackage{graphicx}
\usepackage{cite}
\usepackage{amssymb}

\begin{document}
\title{Revision of the derivation of stellar rates from experiment and impact on Eu s-process contributions}

\author{Thomas Rauscher$^{1,2,3}$}

\address{$^1$ Centre for Astrophysics Research, School of Physics, Astronomy and Mathematics, University of Hertfordshire, Hatfield AL10 9AB, United Kingdom}
\address{$^2$ Institute of Nuclear Research (ATOMKI), H-4001 Debrecen, POB 51, Hungary}
\address{$^3$ Department of Physics, University of Basel, 4056 Basel, Switzerland}

\ead{t.rauscher@herts.ac.uk}

\begin{abstract}
A new, general formalism to include experimental data in revised stellar rates is discussed, containing revised uncertainties. Application to the $s$-process shows that the actual uncertainties in the neutron capture rates can be larger than would be expected from the experimental errors alone. As a specific example for how astrophysical conclusions can depend on the approach selected to derive stellar rates, the $^{151}\mathrm{Eu}/(^{151}\mathrm{Eu}+^{153}\mathrm{Eu})$ abundance ratio from AGB star models is presented. Finally, a recommended workflow for the derivation of stellar rates from experiment is laid out.
\end{abstract}

\section{Introduction}

The estimate of uncertainties in the astrophysical reaction rates included in extended reaction networks is essential for nucleosynthesis studies to disentangle uncertainties stemming from the astrophysical model from those of the nuclear input. Most of the rates included in complete rate libraries are not experimentally constrained and thus prone to theoretical uncertainties which are difficult to assess \cite{sensipaper}. On the other hand, experimental efforts over decades have provided reaction cross sections  (CS) along the line of stability. Measurements of neutron captures for the $s$-process have been especially successful because there is no Coulomb barrier in such reactions which would prevent data to be taken in the astrophysically important energy range. Experiments studying only reactions on the ground state (g.s.) of a nucleus cannot always constrain the stellar rate well, even at $s$-process temperatures \cite{sprocuncert,jphysconf}. In a stellar environment, nuclei are also present in excited states and thus a measurement of the g.s.\ reaction only constrains a fraction of the actual stellar rate. Excited state contributions to stellar rates are especially pronounced for nuclei with low-lying excited states or a high inherent nuclear level density \cite{review}, as found in many nuclei participating in s-process nucleosynthesis. This requires a detailed re-evaluation of the stellar rates and their uncertainties derived from experiment, as it is not possible by default to assign the same uncertainties as the ones derived for the experiment. 

\section{Ground- and excited state contributions to the stellar rate}

\label{sec:defs}

The astrophysical reaction rate $r^*$ for an interaction between two particles or nuclei in a stellar environment is obtained by folding the Maxwell-Boltzmann energy distribution $\Phi$, describing the thermal c.m.\ motion of the interacting nuclei in a plasma of temperature $T$, with a measure $\sigma^*$ of the probability that the reaction occurs and by multiplying the result with the number densities $n_a$, $n_A$, i.e., number of interacting particles in a unit volume,
\begin{equation}
\label{eq:rate}
r^*=\frac{n_a n_A}{1+\delta_{aA}} \int_0^\infty \sigma^*(E) \Phi(E,T)\;dE = \frac{n_a n_A}{1+\delta_{aA}} R^* \quad.
\end{equation}
The stellar reactivity (or rate per particle pair) is denoted by $R^*$. To avoid double counting of pairs, the Kronecker symbol $\delta_{aA}$ is used. Depending on temperature and nuclear level structure, a fraction of nuclei is present in an excited state in the plasma, instead of being in the ground state (g.s.).
The relative contribution $X_i$ of a specific level $i$ to the total stellar rate $r^*$ is given by \cite{sprocuncert}
\begin{equation}
\label{eq:xfactor}
X_i(T)=\frac{2J_i+1}{2J_0+1}e^{-E_i/(kT)}\frac{\int\sigma_i(E)\Phi(E,T)dE}{\int\sigma^\mathrm{eff}(E)\Phi(E,T)dE} \quad.
\end{equation}
 For the g.s.\ ($i=0$), this simplifies to \cite{xfactor}
\begin{equation}
\label{eq:gsxfactor}
X_0(T)=\frac{\int\sigma_0(E)\Phi(E,T)dE}{\int\sigma^\mathrm{eff}(E)\Phi(E,T)dE} \quad.
\end{equation}
The \textit{effective cross section} appearing above is defined as \cite{review,holmes}
\begin{equation}
\sigma^\mathrm{eff}(E) = \sum_i \sum_j \frac{2J_i+1}{2J_0+1} \frac{E-E_i}{E}
\sigma^{i \rightarrow j}(E-E_i) 
\quad,
\label{eq:effcs}
\end{equation}
with
$\sigma_i=\sum_j \sigma^{i \rightarrow j}$ being the sum of partial CS leading to final states $j$.
As usual, CS for individual transitions $\sigma^{i \rightarrow j}$ are zero for negative energies.

Although not discussed in further detail here, it is worth mentioning that also weak interactions and decays are affected by thermal population of excited states and similar quantities can be defined for them in complete analogy.

\section{Implications for the experimental determination of astrophysical reaction rates}
\label{sec:exp}

Laboratory measurements can only determine the g.s.\ CS $\sigma_0^\mathrm{exp}=\sigma_0$ (or the CS of a long-lived isomeric state). This is only sufficient for the determination of the stellar reactivity when the contribution $X_0$ is close to unity. Designing an experiment, it should be taken care to measure a reaction with the largest possible $X_0$. This also implies that the reaction should be measured in the direction of largest $X_0$ \cite{jphysconf,coulsupp}.

When $X_i<1$, only a combination of experimental data and theory can yield a stellar rate. It is not trivial to find the correct combination to derive the stellar rate as well as its new uncertainty after $\sigma_0$ has been measured. Two extreme cases can be found, depending on whether any errors in the prediction of the g.s.\ CS are correlated with the theory errors in the $\sigma_{i>0}$ or not. The first approach is to assume no such correlation. It leads to the largest uncertainties. It should be used whenever a detailed theoretical investigation of the deviations between experiment and theory is missing. Therefore it is the approach of choice for experimentalists or stellar modelers without further support in reaction theory.


Strictly speaking, the experimental CS can only replace one of the contributions to the stellar rate while the others remain unconstrained by the data.
Approach A starts from the notion that only measured quantities should be included in the new stellar rate and its uncertainty estimate, while all other (theoretical) ingredients remain unchanged. Since the stellar reactivity is given by a weighted sum of reactivities
\begin{equation}
R^*_\mathrm{th}=\frac{w_0R_0^\mathrm{th}+w_1R_1^\mathrm{th}+w_2R_2^\mathrm{th}+\dots}{w_0+w_1+w_2+\dots} \quad,
\end{equation}
where $w_i=(2J_i+1)\exp (-E_i/(kT))$ and $J_i$, $E_i$ are spin and excitation energy of the $i$-th excited state, respectively, this implies
\begin{equation}
\label{eq:approach1}
R^*_\mathrm{new,A}=\frac{w_0R_0^\mathrm{exp}+w_1R_1^\mathrm{th}+w_2R_2^\mathrm{th}+\dots}{w_0+w_1+w_2+\dots} \quad.
\end{equation}
Therein, $R_0^\mathrm{exp}$ is the reactivity obtained from folding the experimentally obtained CS on a g.s.\ (or isomeric state) with a Maxwell-Boltzmann distribution, similarly as it is done with the stellar CS $\sigma^*$ in equation (\ref{eq:rate}).
With the experimentally determined reactivity $R_0^\mathrm{exp}$, equation (\ref{eq:approach1}) yields the (temperature dependent) correction factor $f_\mathrm{A}^*(T)$, which has to be applied to the old stellar reactivity to provide the new $R^*_\mathrm{new}(T)=f_\mathrm{A}^*(T) R^*_\mathrm{th}(T)$,
\begin{equation}
f_\mathrm{A}^*(T)=1+X_0(T)\left(\frac{R_0^\mathrm{exp}(T)}{R_0^\mathrm{th}(T)}-1\right) \label{eq:renormstellar} \quad,
\end{equation}
containing the ratio between experimental and theoretical g.s.\ reactivity.

It is important to note that the determination of the rate is closely entwined with its associated uncertainty or ``error'' and should never be quoted and used without it. Frequently used in astrophysics are
uncertainty factors $U$, which assume that the ``true'' value $R^*_\mathrm{true}$ of a (semi-)theoretical reactivity $R^*$ is in the range $UR^*\geq R^*_\mathrm{true} \geq R^*/U$. It is neither trivial to estimate the theory uncertainty nor the experimental one, especially when it has to be in a consistent, intercomparable manner. The fundamental differences between error determinations in experiment and theory are discussed in \cite{sensipaper} and appropriate choices are suggested in \cite{sprocuncert}. Here, it should just be mentioned that, unlike experimental measurements, theory results cannot usually be assumed to be randomly drawn from a defined statistical distribution. A theoretical model or parametrization can simply be inadequate and it is impossible to define a set of statistically distributed models. Thus, the theory uncertainty is more closely related to a systematic error and equally difficult to estimate.

Once having estimated the uncertainties in some way, nevertheless, in the above approach A it is simple to calculate the new uncertainty factor of the stellar rate from a combination of the uncertainty factor $U^*_{\mathrm{th}}=U_0^\mathrm{th}$ of the theoretical stellar reactivity $R^*$ and an experimental uncertainty $U_{\mathrm{exp}}=U_0^\mathrm{exp}\leq U^*_{\mathrm{th}}$ of the measured g.s.\ contribution\footnote{For simplicity, $U^*_{\mathrm{th}}=U_0^\mathrm{th}$ is assumed here but a new uncertainty can also be derived rigorously when using different uncertainties for each $R_i^\mathrm{th}$. It is further possible to consider the uncertainty of $X_0$ in $U^*_\mathrm{new}$. Its impact, however,
is small with respect to the other experimental and theoretical uncertainties \cite{xfactor}.}.  This yields the uncertainty factor of the new stellar rate
\begin{equation}
U^*_\mathrm{new,A}(T)=U_{\mathrm{exp}}(T)+(U^*_{\mathrm{th}}(T)-U_{\mathrm{exp}}(T))(1-X_0(T)) \label{eq:uncertainty} \quad.
\end{equation}


As pointed out in \cite{sprocuncert}, the other extreme approach would be to assume that all theoretical uncertainties, i.e., also those of the reactions proceeding on excited states, are removed once the g.s.\ reactions and their deviation from the predictions have been determined experimentally. This implies two strong assumptions on theory: (a) the cause of any discrepancy between $R_0^\mathrm{th}$ and $R_0^\mathrm{exp}$ also causes a similar deviation of the same magnitude in all $R_{i>0}^\mathrm{th}$, and (b) there are no further uncertainties in predicted excited state CS. Only when both assumptions are valid, the same renormalization can be applied to g.s.\ and excited state transitions, i.e.,
\begin{eqnarray}
R^*_\mathrm{new,B}&=&\frac{w_0R_0^\mathrm{exp}+w_1R_1^\mathrm{th}\frac{R_0^\mathrm{exp}}{R_0^\mathrm{th}}+w_2R_2^\mathrm{th}\frac{R_0^\mathrm{exp}}{R_0^\mathrm{th}}+\dots}{w_0+w_1+w_2+\dots} = \nonumber \\
& =& \frac{R_0^\mathrm{exp}}{R_0^\mathrm{th}} \times \frac{w_0R_0^\mathrm{th}+w_1R_1^\mathrm{th}+w_2R_2^\mathrm{th}+\dots}{w_0+w_1+w_2+\dots} \quad.
\end{eqnarray}
In this approach B, therefore, the renormalization factor $f_\mathrm{B}^*$ for the stellar reactivity (and rate) is $f_\mathrm{B}^*=R_0^\mathrm{exp}/R_0^\mathrm{th}$, so that $R^*_\mathrm{new,B}=f_\mathrm{B}^* R^*_\mathrm{th}$, as before. It has to be noted that this is equivalent to multiplying the measured g.s.\ reactivity $R_0^\mathrm{exp}$ by the so-called \textit{stellar enhancement factor} (SEF) $f_\mathrm{SEF}=R^*_\mathrm{th}/R_0^\mathrm{exp}$ because of assumption (a). Until recently this was the favored approach, even when nothing was known about the causes of the deviations between prediction and experiment and also not whether these also apply similarly to reactions from the excited states \cite{bao,starlib}.

The uncertainty for the new stellar rate obtained in approach B is difficult to quantify. Only if both assumptions (a) and (b) really apply, and only then, it is just the uncertainty of the measurement. If this is not the case, the uncertainty will be larger. In order to be on the safe side, it may be advisable to use the error estimate $U^*_\mathrm{new,A}$ from equation (\ref{eq:uncertainty}) even when using approach B to derive $R^*_\mathrm{new,B}$. 

The above two approaches are the two extreme cases. Realistically, the actual stellar rate and its uncertainty may be anywhere between these two cases. Detailed theoretical investigations have to be performed separately for each nucleus and each reaction before conclusions can be drawn from the comparison of the experimental results and the prediction. In the absence of such a detailed theoretical investigation, it is preferrable to apply the ``pessimistic view'' of approach A, contrary to the historical custom of using the seemingly simpler approach B. Although approach A will lead to larger uncertainties, it encompasses the values of approach B and thus covers all possibilities.

There are a number of reasons why the uncertainties in the predictions of the transitions originating on excited states can be different than those of the ones commencing on the g.s.\ of the target nucleus. First, it is clearly seen in equation (\ref{eq:effcs}) that $\sigma^\mathrm{eff}$ includes transitions at a range of relative interaction energies $0\leq E-E_i\leq E_\mathrm{G}-E_i$, where $E_\mathrm{G}$ is the upper end of the relevant energy window at each temperature. It is known that the sensitivity of the $\sigma^{i \rightarrow j}$ are strongly energy-dependent (see, e.g., \cite{sensipaper}) and therefore transitions from excited states (occurring at lower relative energy) may be sensitive to different nuclear properties than those from the g.s. For example, for neutron captures in the $s$-process this can be important mainly for nuclei with large capture $Q$-value and high nuclear level densities, for which the neutron widths may become comparable to or smaller than the $\gamma$-widths within the covered energy range. Second, even if the sensitivities to different widths are not changing, different spins and parities of the excited states imply different angular momenta and therefore different angular momentum barriers in particle transitions, and also may give rise to a different selection of electromagnetic multipolarities in $\gamma$-transitions. This may be more important in nuclei with low level densities. The actual circumstances and sensitivities will be different in each application and have to be thoroughly investigated for each reaction separately.

\section{Impact on the Eu production in the s-process}

\begin{figure}
\begin{center}
\includegraphics[width=0.6\textwidth]{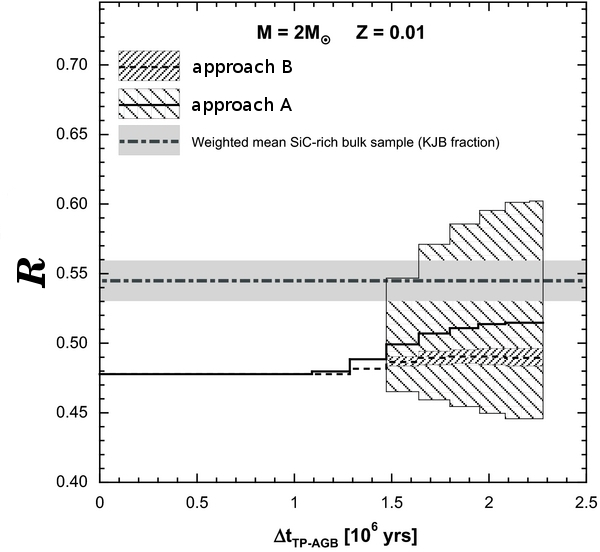}
\caption{Evolution of $\mathcal{R}$ in the AGB envelope (time
from the beginning of the TP-AGB phase) for two models
using the same experimental values \cite{marrone} for $^{151}$Sm(n,$\gamma$), but the two approaches discussed in the text. The shaded area represents the range of results obtained when varying the rate with the derived uncertainty factors. [This is a modified version of Fig.\ 4 in \cite{avila}.] \label{fig:eu}}
\end{center}
\end{figure}

An illustrative example for the importance of choosing the right approach to derive the stellar rate from experiment is the reproduction of the abundance ratio $\mathcal{R}=Y_{^{151}\mathrm{Eu}}/Y_\mathrm{Eu}$ found in meteoritic material \cite{avila} and in CEMP stars \cite{aoki}. Recently, individual mainstream stardust silicon carbide (SiC) grains and a SiC-enriched bulk sample (KJB) from the Murchison
carbonaceous meteorite have been analyzed for Eu isotopes \cite{avila}. The mainstream grains are believed to have condensed in the outflows of $\approx 1.5-3$ $M_\odot$ carbon-rich asymptotic giant branch (AGB) stars with close-to-solar metallicity and $\mathcal{R}$ is determined by the values of the $^{151}$Sm(n,$\gamma$) rate during $s$-processing. There are also abundances available from previous astronomical observations
of carbon-enhanced metal-poor stars enriched in $s$-process elements \cite{aoki}. Despite
the difference in metallicity between the parent stars of the grains and the metal-poor stars, the $\mathcal{R}$ values
derived from the meteoritic data agree well with those derived from astronomical observations. They appear systematically higher than the solar value of 0.48 but show large error bars. The new data from the SiC aggregate KJB, however, provides a more precise result ($\mathcal{R} = 0.54 \pm 0.03$, 95\% conf.), which is compatible with both the mainstream and the stellar data. It clearly shows an enhancement with respect to the solar value. 

Sensitivity studies find that $^{151}$Sm(n,$\gamma$)$^{152}$Sm is the crucial reaction determining $\mathcal{R}$ \cite{avila}. Although the stellar $\beta$-decay rate of $^{151}$Sm is only theoretically known, this neutron capture reaction dominates the uncertainties \cite{avila}. A recent measurement of $^{151}$Sm(n,$\gamma$) constrained $R_0^\mathrm{exp}$ within a few percent \cite{marrone}. Current $s$-process predictions using modern AGB models and including these (n,$\gamma$) data, however, cannot reproduce the enhancement in $\mathcal{R}$ when choosing the traditional approach B to derive the stellar rate. Figure \ref{fig:eu} shows that, on the other hand, the results obtained with a rate derived with approach A are consistent with the meteoritic ratio.

It is sometimes stated that the $s$-process is the nucleosynthesis process with the best constrained nuclear input (see, e.g., \cite{kae11,sreview}). This statement, however, is based on the uncertainties in the measured (n,$\gamma$) CS. The resulting uncertainties $U^*_\mathrm{new}$ in the stellar reactivities and rates may be larger in many cases \cite{sprocuncert}.

\section{Conclusion}

The revised approach to improve stellar rates with experimental CS presented here allows for the first time to also consistently discuss how rate uncertainties are affected by the inclusion of experimental data. In intermediate and heavy nuclei, excited state contributions to the stellar rate are not negligible even at the comparatively low $s$-process temperatures and limit the ability of reaction measurements to constrain the stellar rates. This is even more pronounced in charged-particle reactions as well as in nucleosynthesis processes at higher temperature.

If nothing else is known, approach A as described above is the recommended method to derive the stellar rate und its uncertainty.
Nevertheless, the best way to derive stellar rates from experiment is to perform a detailed theoretical study of all transitions and their sensitivities to nuclear input which contribute in the reaction in question. Only such a study allows to judge how well the excited state CS are constrained by a measurement of the g.s.\ CS.

The preferred workflow after having measured $\sigma_0^\mathrm{exp}$ across the range of astrophysically relevant energies, therefore, would be the following:
\begin{enumerate}
\item Check whether $X_0(T)$ deviate from unity within the astrophysical temperature range; if $X_0 \approx 1$, the stellar rate can be directly calculated from the measured CS and the uncertainty is given by the experimental one. ($X_0$ can be found in \cite{sensipaper,sprocuncert,online}.)
\item For $X_0<1$, if no further theory work is available, use approach A, i.e., determine the new stellar rate using $f_\mathrm{A}^*$ from equation (\ref{eq:renormstellar}) and its uncertainty from equation (\ref{eq:uncertainty}). The required $R_0^\mathrm{th}$ can be, e.g., found in \cite{holmes,online} but have to be consistent with the renormalized $R^*_\mathrm{th}$.
\item For $X_0<1$, if theory work is desired, the transitions on the g.s.\ and excited states and their sensitivities have to be studied in detail:
\begin{enumerate}
\item If discrepancies between g.s.\ prediction and experiment are found to apply similarly to the excited state transition, use  $f_\mathrm{B}^*$ to derive the stellar rate.
\begin{enumerate}
\item If there are no remaining uncertainties (from other sources) in the predicted transitions from excited states, the final uncertainty of the stellar rate is the experimental one.
\item If there are further uncertainties but no theory insights on their scaling, use equation (\ref{eq:uncertainty}) for the uncertainty of the stellar rate.
\item Otherwise, theory should attempt to find an ``intermediate'' uncertainty factor, i.e., with a value between those obtained with approaches A and B.
\end{enumerate}
\item If the theoretical excited state CS cannot be scaled in the same way as the g.s.\ CS:
\begin{enumerate}
\item If there are no further theory insights, use approach A for both stellar rate (from equation \ref{eq:renormstellar}) and its uncertainty (equation \ref{eq:uncertainty}).
\item Otherwise, theory should attempt to find ``intermediate'' renormalization and uncertainty factors, i.e., with values between those obtained with approaches A and B.
\end{enumerate}
\end{enumerate}
\end{enumerate}

\ack
This work was supported in part by the Swiss NSF, the European Commission within the FP7 ENSAR/THEXO project, and the EuroGENESIS Collaborative Research Programme. TR also acknowledges support through a "Distinguished Guest Scientist Fellowship" from the Hungarian Academy of Sciences.


\section*{References}
\begin{thebibliography}{99}
\bibitem{sensipaper} Rauscher T 2012 \textit{Ap.\ J. Suppl.} \textbf{201} 26
\bibitem{sprocuncert} Rauscher T 2012 \textit{Ap.\ J. Lett.} \textbf{755} L10
\bibitem{jphysconf} Rauscher T 2013 \textit{J. Phys.\ Conf.\ Ser.} \textbf{420} 012138
\bibitem{review} Rauscher T 2011 \textit{Int.\ J. Mod.\ Phys.} E \textbf{20} 1071
\bibitem{xfactor} Rauscher T, Mohr P, Dillmann I and Plag R 2011 \textit{Ap.\ J.} \textbf{738} 143
\bibitem{holmes} Holmes J A, Woosley S E, Fowler W A and Zimmerman B A 1976 \textit{At.\ Data Nucl.\ Data Tables} \textbf{18} 305
\bibitem{coulsupp} Rauscher T, Kiss G G, Gy\"urky Gy, Simon A, F\"ul\"op Zs and Somorjai E 2009 \textit{Phys.\ Rev.} C \textbf{80} 035801
\bibitem{bao} Bao Z Y, Beer H, K\"appeler F, Voss F, Wisshak K and Rauscher T 2000 \textit{At.\ Data Nucl.\ Data Tables} \textbf{76} 70 
\bibitem{starlib} Sallaska A L, Iliadis C, Champagne A E, Goriely S, Starrfield S and Timmes F X 2013 \textit{Ap.\ J. Suppl.} \textbf{207} 18
\bibitem{avila} \'Avila J N, et al 2013 \textit{Ap.\ J. Lett.} \textbf{768} L18
\bibitem{aoki} Aoki W, et al 2003 \textit{Ap.\ J. Lett.} \textbf{592} L67
\bibitem{marrone} Marrone S, et al 2006 \textit{Phys.\ Rev.} C \textbf{73} 034604
\bibitem{kae11} K\"appeler F,  Gallino R, Bisterzo S and Aoki W 2011 \textit{Rev.\ Mod.\ Phys.} \textbf{83} 157
\bibitem{sreview} Wiescher M, K\"appeler F and Langanke K 2012 \textit{Ann.\ Rev.\ Astron.\ Astrophys.} \textbf{50} 165
\bibitem{online} Online interface to the NON-SMOKER cross section and rate library, http://nucastro.org/nonsmoker.html
\end{thebibliography}
\end{document}